\documentclass[aps,prl,twocolumn]{revtex4}
\usepackage[english]{babel}
\usepackage[dvips]{graphics,color}
\usepackage{newlfont,epsfig}
\usepackage{graphicx}
\usepackage{amsmath}
\usepackage{amssymb}
\usepackage{bm}
\newcommand{\ket}[1]{\ensuremath{\left|{#1}\right\rangle}}

\begin{document}
\title{Generation of a two-photon singlet beam}
\author{W. A. T. Nogueira}
\email{wallon@fisica.ufmg.br}
\affiliation{Universidade Federal de
Minas Gerais, Caixa Postal 702, Belo Horizonte, MG 30123-970,
Brazil}
\author{S. P. Walborn}
\affiliation{Universidade Federal de Minas Gerais, Caixa Postal
702, Belo Horizonte, MG 30123-970, Brazil}
\author{S. P\'adua}
\affiliation{Universidade Federal de Minas Gerais, Caixa Postal
702, Belo Horizonte, MG 30123-970, Brazil}
\author{C. H. Monken}
\affiliation{Universidade Federal de Minas Gerais, Caixa Postal
702, Belo Horizonte, MG 30123-970, Brazil}
\date{\today}
\begin{abstract}
Controlling the pump beam transverse profile in multimode
Hong-Ou-Mandel interference, we generate a ``localized" two-photon
singlet state, in which both photons propagate in the same beam.
This type of multi-photon singlet beam may useful in quantum
communication to avoid decoherence. We show that although the
photons are part of the same beam, they are never in the same
plane wave mode, which  is characterized by spatial antibunching
behavior in the plane normal to the propagation direction.
\end{abstract}
 \maketitle Entangled multi-photon
polarization states are an important tool in the investigation and
future implementation of quantum information protocols
\cite{Nielsen}.  In the case of the polarization of two photons,
the maximally-entangled Bell states, given by
\begin{subequations}
    \label{eq:1}
    \begin{align}
    \langle\psi^{\pm}\rangle & = \frac{1}{\sqrt{2}}(\langle H \rangle_{1}\langle V \rangle_{2}
    \pm \langle V \rangle_{1}\langle H \rangle_{2} )  \\
    \langle \phi^{\pm} \rangle & = \frac{1}{\sqrt{2}}(\langle H \rangle_{1}\langle H \rangle_{2}
    \pm \langle V \rangle_{1}\langle V \rangle_{2} )
    \end{align}
\end{subequations}
form a complete basis in four-dimensional Hilbert space.  Here $H$
and $V$ are horizontal and vertical polarization and kets $1$ and
$2$ represent plane-wave modes. The ``triplet" states,
$\ket{\psi^{+}}$ and $\ket{\phi^{\pm}}$, are symmetric and the
``singlet" state $\ket{\psi^{-}}$ is antisymmetric under exchange
of the two photons.  To maintain their overall bosonic symmetry,
photons in the singlet polarization state also display spatial
antisymmetry, and cannot occupy the same plane-wave mode
\cite{Zeilinger1,Zeilinger2}.  This behavior can be seen in the
usual two-photon interference experiment:   when two
indistinguishable plane-wave photons meet at a beam splitter (BS),
they leave the BS in the same port if they are in a symmetric
polarization state and in opposite ports if they are in the
antisymmetric $\ket{\psi^{-}}$ polarization state.
\par
The antisymmetry exhibited by the singlet state leads to some
interesting properties.  Recently, some attention has been paid to
``supersinglet" states \cite{Cabello1} -- singlet states of two or
more particles. It has been shown that these states can be used to
solve several problems that have no classical solutions, as well
as in violations of Bell-type inequalities and in proofs of Bell's
theorem without inequalities \footnote{See \cite{Cabello1} for a
review.}. Perhaps an even bigger potential is in the storage and
transmission of quantum information. In particular, singlet states
$\ket{\psi_{N}^{-}}$ formed by $N$ two-dimensional systems --
qubits -- can be used to construct decoherence-free subspaces
which are robust to collective decoherence
\cite{Dfs,Zanardi1,Cabello1}, in which the system--environment
interaction is the same for all qubits.  More specifically, these
states are (up to a global phase factor) invariant to any type of
$N$-lateral unitary operation,
\begin{equation}\label{eq:U}
  U^{\otimes N} \langle \psi_{N}^{-} \rangle = \langle \psi_{N}^{-} \rangle,
\end{equation}
where $U$ is a single qubit unitary operation and $U^{\otimes N}$
is given by $U \otimes U \otimes \ldots \otimes U$. Hence, it is
possible to avoid collective decoherence of this form by encoding
quantum information in the $\ket{\psi_{N}^{-}}$ states
\cite{Zanardi1}.  The assumption that the decoherence is
collective is generally valid as long as the physical systems
representing the qubits are closely spaced as compared to the
coherence length of the environment \cite{Cabello1}.   In future
implementations of optical quantum communication, for example,
this may not be true if the photons are not propagating in the
same spatio-temporal region.
\par
It has been shown experimentally that the two-photon state
$\ket{\psi^{-}} \equiv \ket{\psi_{2}^{-}}$ is robust to
decoherence of the form (\ref{eq:U}) \cite{Kwiat1}. Photons in the
polarization state $| \psi^{-} \rangle$ were generated using
spontaneous parametric down-conversion (SPDC). Each photon was
subject to a decohering environment in the form of a birefringent
crystal, which introduces a frequency-dependent random phase
between horizontal and vertical polarization components. To
simulate a collective environment, the crystals were kept aligned
so that both photons always suffered the same decoherence.  In
this way, it was shown that the fidelity of the $\ket{\psi^{-}}$
state is unaffected by the decohering crystals.
\par
We could assure that the decoherence suffered by the
$|\psi^{-}\rangle$ state is more likely to be collective if we
could localize the two-photons to within a given spatio-temporal
region, such as a well-collimated beam for example.  Here we show
experimentally that, using multimode Hong-Ou-Mandel (HOM)
interference \cite{Walborn1}, it is possible to create a
``localized" $|\psi^{-} \rangle$ polarization state, in which the
two photons propagate in a single beam. In this manner, up to a
scale defined by the beam width, any unitary decoherence caused by
the environment is felt equally by the two photons.
\par
\begin{figure}
{\includegraphics[width=7cm]{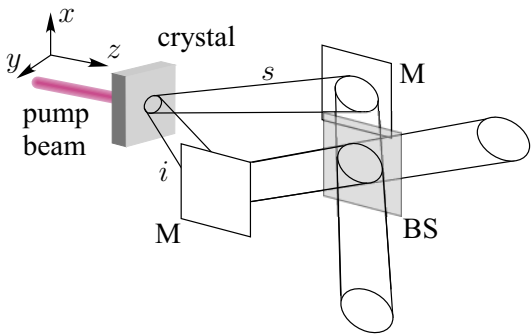}}
\protect\caption{\label{HOMesquema} HOM interferometer. Two
photons created by non-collinear SPDC are directed by mirrors (M)
onto a non-polarizing 50-50 beam splitter (BS).  Path lengths $s$
and $i$ can be made equal by the translation of one of the mirrors
M.}
\end{figure}

For years HOM interferometry \cite{Hong1} (and variations) has
been one of the principal methods used to observe two-photon
interference. As shown in Fig. \ref{HOMesquema}, two photons $s$
and $i$ are created by non-collinear SPDC and directed onto a
non-polarizing 50-50 beam splitter (BS). If the optical path
lengths of $s$ and $i$ are equal, then they interfere as described
above.
\par
Recently, Walborn \textit{et. al.} \cite{Walborn1} showed that in
a multimode treatment of HOM interference,it is necessary to take
into account both the polarization and transverse spatial degrees
of freedom. In multimode non-collinear SPDC, it is well known
that, under certain experimental conditions, the transverse
profile of the pump beam field $\mathcal{W}(x,y,z)$ is transferred
to the two-photon detection amplitude as
$\mathcal{W}((x_{1}+x_{2})/2,(y_{1}+y_{2})/2,Z)$  \cite{Monken1}.
In the monochromatic approximation considered here (it is assumed
that the down-converted photons have the same wavelength), the
two-photon detection amplitude can be regarded as the two-photon
wave function \cite{Mandel1}.  Subjecting the down-converted
photons to a beam splitter, the observed HOM interference then
depends upon the parity of the function $\mathcal{W}(x,y,z)$.
Specifically, using a pump beam that is an odd function of the $y$
coordinate, $\mathcal{W}(x,-y,z) = -\mathcal{W}(x,y,z)$, photons
in the polarization state $\langle \psi^{-} \rangle$ leave the
beam splitter in the same output port.  In this case, following
\cite{Walborn1},
 the probability
amplitude to detect \emph{both} photons in the \emph{same} output port is
given by
\begin{align}\label{amplit1}
\bm{\Psi}(\mathbf{r}_{1},\mathbf{r}_{2}) \propto
 & \, \mathcal{W} \left(\frac{x_{1}+x_{2}}{2},\frac{y_{1}-y_{2}}{2},Z \right) \nonumber \\
& \times  (\bm{H}_{1} \bm{V}_{2}- \bm{V}_{1} \bm{H}_{2})
\end{align}
where $\mathbf{r}_{1} = (x_{1},y_{1},z_{1})$ and $\mathbf{r}_{2} =
(x_{2},y_{2},z_{2})$ are the coordinates of detectors $D_{1}$ and
$D_{2}$, respectively, with $z_{1} = z_{2} = Z$.  We note that
both detectors are placed in the same output port of the BS,
\textit{i.e.}, $D_{1}$ and $D_{2}$ detect in the same spatial
region. $\bm{H}$ and $\bm{V}$ are unit polarization vectors in the
$H$ and $V$ directions.  The $y_{1}-y_{2}$ dependence of
(\ref{amplit1}) is due to the reflection of one of the photons at
the beam splitter \cite{Walborn1}.  Here it has been assumed that
the 50-50 beam splitter is symmetric.  In addition, we have
ignored the entanglement between the polarization and wave vector
due to the birefringence of the nonlinear crystal, which can be
minimized using a compensating crystal in addition to narrow band
interference filters and small detection apertures in the
experimental setup.
\par
In contrast to the experiment reported in ref. \cite{Walborn1},
here we focus our attention on the polarization and spatial
properties of the two-photon beam that comes out of one of the
beam splitter ports. Although photons in the $\langle \psi^{-}
\rangle$ always leave through the same port, which port they leave
through is random. We also note that both the spatial and
polarization components of Eq. (\ref{amplit1}) are antisymmetric.
 \begin{figure}
\includegraphics[width=7cm]{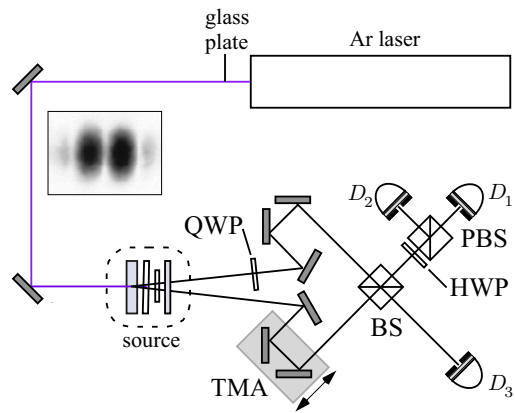}
 \caption{\label{fig:setup} Experimental setup. A glass plate is
placed halfway into the beam and adjusted to create $\pi$ phase
difference between the two halves, creating a profile that is an
odd function of the $y$ coordinate.  The inset shows a photograph
of the pump profile in the detection region. A 2-mm-long nonlinear
crystal (BBO) is pumped by an argon laser beam generating twin
photons in crossed cones. The ``source'' in figure is composed of
the nonlinear crystal, 1 mm compensating crystal, UV filter and
half wave plate as in  \cite{Kwiat2}. QWP is a quarter wave plate
used to change the state from $| \psi^{+} \rangle$ to $| \psi^{-}
\rangle$. BS is a 50/50 beam splitter. The trombone mirror
assembly (TMA), mounted on a computer-controlled motorized stage,
is used to adjust the path length difference.  The polarizing beam
splitter (PBS) and half-wave plate (HWP) are used to detect
photons in the same output of the BS.  $D_{1}$, $D_{2}$ and
$D_{3}$ are photodetectors.}
\end{figure}
\par
Fig. \ref{fig:setup} shows the experimental setup. At the output
of the argon laser ($\lambda_{p}=351$ nm), we inserted a thin
($\sim 150\,\mu$m) glass laminate halfway into the Gaussian
profile pump beam and adjusted the angle in order to achieve a
$\pi$ phase difference between the two halves of the beam. This
produces a transverse profile that is an odd function of the
horizontal $y$ coordinate. A photograph of the beam intensity
profile in the detection region ( $\sim 3$\,m from the glass
laminate) is shown in  the inset of Fig. \ref{fig:setup}. Because
of the spatial filtering due to propagation of the beam, this
profile presents just one central minimum. In the far-field
region, this beam is similar to the first-order Hermite-Gaussian
beam HG$_{01}$.
\par
This beam is used to pump a 2-mm thick nonlinear crystal (BBO) cut
for degenerate type II phase matching. The crystal is adjusted to
generate polarization-entangled photons ($\lambda \sim 702$ nm)
using the crossed-cone source as reported in \cite{Kwiat2}.  The
output state of this source is controlled by adjusting the angle
of the compensating crystal to be the $| \psi^{+} \rangle$
polarization state. With a quarter-wave plate (QWP) in one of the
paths, the relative phase can be manipulated in order to change
from the polarization state $| \psi^{+} \rangle$ to $| \psi^{-}
\rangle$ \cite{Kwiat2}.  A trombone mirror assembly (TMA) is
mounted on a motorized translational stage to adjust the
path-length difference of the interferometer.  The photons are
directed onto a beam splitter (BS).  $D_{1}$, $D_{2}$, and $D_{3}$
are EG\&G SPCM 200 photodetectors equipped with interference
filters ($1$\,nm FWHM centered at $702$\,nm) and $3$ mm circular
detection apertures.  A computer was used to register coincidence
and single counts.
\par
With the BS removed, we used polarization analyzers (not shown in
Fig. \ref{fig:setup}) consisting of a half-wave plate (HWP) and
polarizing beam splitter (PBS) to test the quality of the $\langle
\psi^{-} \rangle$ polarization state generated by the crystal. We
did this by observing the usual polarization interference
\cite{Kwiat2}: one polarizer was kept fixed at $0^{\circ}$ or
$45^{\circ}$ while the other was rotated. We observed interference
curves with visibilities greater than $0.97 \pm 0.01$ in both
cases, implying a high degree of polarization entanglement,
\textit{i.e.}, a high quality $| \psi^{-} \rangle$ state. \par
Putting the BS in place and removing the polarization analyzers,
we measured the usual HOM interference curve in coincidence
detections at the output ports (detectors $D_{1}$ and $D_{3}$ in
Fig. \ref{fig:setup}) by scanning the TMA. With the glass plate
removed we observed interference curves with visibilities
$\mathcal{V}_{HOM} = 0.92\pm0.01$, indicating good spatial overlap
at the BS. With the glass plate placed in the laser beam (odd pump
profile) and using photons in the $\langle \psi^{-} \rangle$
polarization state, the visibility was $\mathcal{V}_{HOM} =
0.82\pm0.01$.  The decrease in visibility was most likely due to
two reasons.  First, the alignment of the HOM interferometer is
noticeable more sensitive when an odd pump beam is used
\cite{Walborn1}.  Second, there is a slight loss in the intensity
of the portion of the pump beam that passes through the glass
laminate,
which creates a small distinguishability in the fourth-order interference.    %
\begin{figure}\includegraphics[width=7cm]{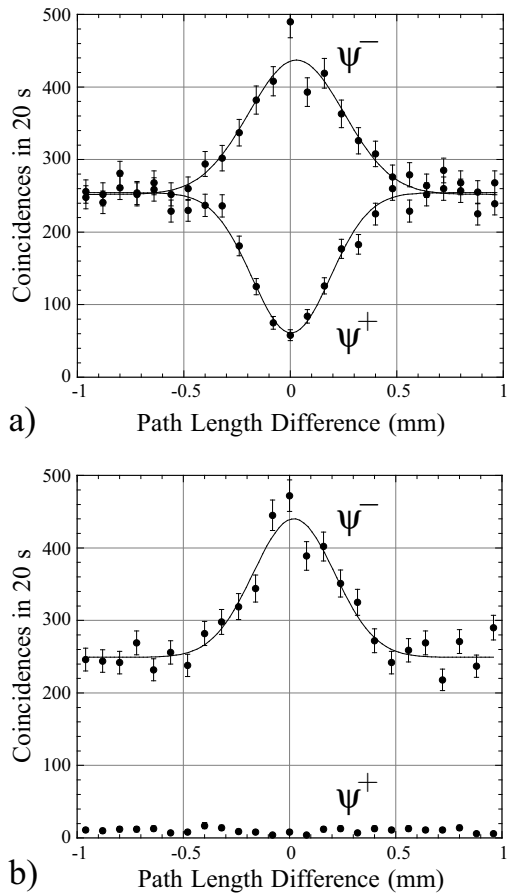}
\protect\caption{\label{fig:4}Detections in the same output port of the BS.
a) Detections in the $H/V$ basis. The
visibility of the state $| \psi{-} \rangle$ is $\mathcal{V} = 0.73
\pm 0.05$. The visibility of the state $| \psi{+} \rangle$ is
$\mathcal{V} = 0.76 \pm 0.02$. b) Detections in the
$+/-$ basis. The visibility of the state $| \psi{-} \rangle$ is
$\mathcal{V} = 0.76 \pm 0.05$. There are no coincidences of the state $|
\psi{+} \rangle$ in this basis.}
\end{figure}
\par
Next, we placed a polarization analyzer (a HWP and PBS) in one
output of the  BS and detected coincidences at the two output
ports of the PBS, so that detectors $D_{1}$ and $D_{2}$ (Fig.
\ref{fig:setup}) always detect orthogonal polarizations. The HWP
was set so the analyzer detected in the $H/V$ or $+/-$ bases,
where $\pm = 1/\sqrt{2}(H \pm V)$.   We scanned the path length
difference and performed HOM interference measurements, however,
this time coincidences were registered at detectors $D_{1}$ and
$D_{2}$. The results are shown in Fig. \ref{fig:4}.  Error bars in
all figures correspond to photon counting statistics
\cite{Mandel1}. Using the $\langle \psi^{-} \rangle$ state, we
observe constructive interference at detectors $D_{1}$ and $D_{2}$
in both the $H/V$ (Fig. \ref{fig:4}a) and $+/-$ bases (Fig.
\ref{fig:4}b). Observing constructive interference in both
detection bases is characteristic of the $\langle \psi^{-}
\rangle$ state, since it is the only antisymmetric  two-photon
polarization state and is invariant to bilateral rotation
\cite{Cabello1}.  In this respect, one can regard the HWP as a
special case of a decoherence environment. Comparatively, using
the polarization state $| \psi^{+} \rangle$, and detecting in the
$H/V$ basis, we observe an interference ``dip" (Fig.
\ref{fig:4}a). However, in the $+/-$ basis (Fig. \ref{fig:4}b), we
observe no coincidences, since in this basis
 the $| \psi^{+} \rangle$ state is proportional to $(\langle + \rangle_{1}
\langle + \rangle_{2} - \langle - \rangle_{1} \langle - \rangle_{2})$. %
\begin{figure}\includegraphics[width=7cm]{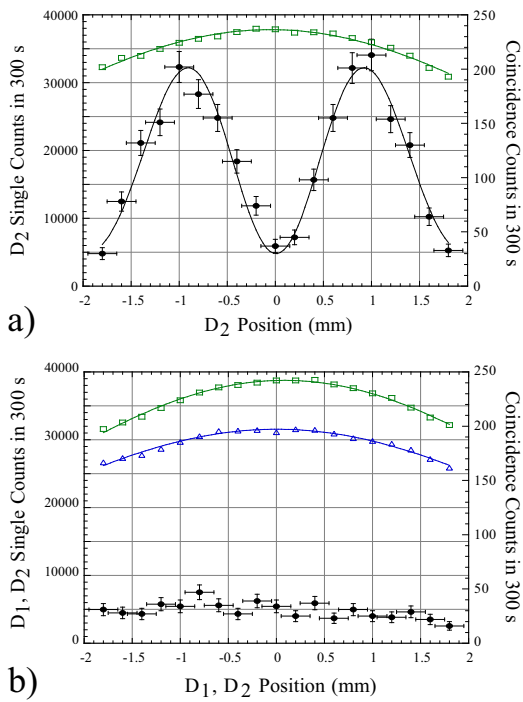}
\protect\caption{\label{fig:5} $\bigtriangleup = D_{1}$ single
counts, $\Box = D_{2}$ single counts, $\bullet =$ Coincidences.
Horizontal error bars correspond to the width of detection slits.
a) Coincidence counts with $D_{1}$ fixed at ``0" and $D_{2}$
scanned horizontally. b) $D_{1}$ and $D_{2}$ scanned together,
always detecting in the same position.}
\end{figure}
\par
It is interesting to examine this experiment from the point of
view of symmetry. In order for the wave packets of the two twin
photons  to occupy the same spatio-temporal region, the total
biphoton wave function must be symmetric. In this symmetrization,
all degrees of freedom must be considered. In the case of the
$\langle \psi^{-} \rangle$ polarization state with an odd pump
beam profile, overall bosonic symmetry requires that photons pairs
are found in the same output port of the BS. However, because of
the antisymmetry of the transverse spatial component, which is
provided by the odd pump beam together with the reflection of one
photon at the beam splitter, the photons are spatially separated
in the $y$-direction and thus do not occupy the same plane wave
mode.  Interestingly enough, this characteristic guarantees that
the singlet beam exhibits spatial antibunching, a quantum effect
with no classical analog \cite{Nogueira1,Nogueira2}.
\par
To investigate this aspect of the singlet beam, both detectors
were equipped with $0.3$ mm $\times \, 3$ mm vertical detection
slits and aligned to the same spatial region as in
\cite{Nogueira1}.  The TMA was set at the interference maximum
(``0" in Fig. \ref{fig:4}). Fig. \ref{fig:5}a shows the
coincidence counts when detector $D_{1}$ is fixed at ``$0$" and
$D_{2}$ is scanned in the horizontal $y$ direction. There is a
coincidence minimum at the origin where $D_{1}$ and $D_{2}$ are
detecting at the same position. The solid line is a curve fit as
in \cite{Nogueira1}. Fig. \ref{fig:5}b shows results when the two
detectors are scanned together in the same sense -- always
detecting in the same position -- in which they always detect a
coincidence minimum. The residual coincidence detections at the
minima are due to the width of the detection slits. In reference
\cite{Nogueira1}, the two photons were in a singlet polarization
state after the birefringent double-slit, but they did not
constitute a beam. The measurements shown above (Fig.
\ref{fig:5}), however, are not of a fourth-order interference
pattern that exhibits spatial antibunching in a detection region,
but rather measurements of the transverse profile of a two-photon
spatially antibunched singlet beam. It is worth noting that while
the singlet beam is necessarily spatially antibunched, spatial
antibunching can also be achieved with symmetric polarization
states and an even pump beam \cite{Ribeiro1}.
\par
Although the photons never occupy the same plane wave mode, it is
important to stress that the individual photons are
indistinguishable in all degrees of freedom (spatially,
temporally, polarization, frequency, etc.). This guarantees that
the decoherence felt by this type of localized state is collective
up to the width of the two-photon beam.
\par
Here we have taken a first step in creating a localized
multi-photon state that is more resistant to decoherence. Using
multimode Hong-Ou-Mandel interference, we have generated a
two-photon singlet beam, which forms a unidimensional decoherence
free subspace. We expect to use these same techniques to create
singlet beams of more than two photons, which could be used to
encode and transmit quantum information in a higher-dimensional
decoherence-free channel \footnote{Experimental work in this
direction is currently being conducted in our laboratory.}.  We
have also shown that the singlet beam is inherently non-classical,
exhibiting spatial antibunching in the transverse plane.

\begin{acknowledgments}

The authors acknowledge financial support from the Brazilian
agencies CNPq and  CAPES.

\end{acknowledgments}

\end{document}